\documentclass[sigconf]{acmart}

\copyrightyear{2018}
\acmYear{2018}
\setcopyright{rightsretained}
\acmConference[CCS '18]{2018 ACM SIGSAC Conference on Computer \& Communications Security}{October 15--19, 2018}{Toronto, ON, Canada}
\acmPrice{}
\acmDOI{10.1145/3243734.3278496}
\acmISBN{978-1-4503-5693-0/18/10}

\author{Gili Rusak, Abdullah Al-Dujaili, Una-May O'Reilly}
	\affiliation{
	\institution{CSAIL, MIT, USA}}
\email{gili@stanford.edu, aldujail@mit.edu, unamay@csail.mit.edu}

\usepackage{amsmath}
\usepackage{xspace}

\usepackage{todonotes}
\usepackage{paralist}
\DeclareMathOperator{\vc}{vec}

\begin{document}
\title{POSTER: AST-Based Deep Learning for Detecting \\ Malicious PowerShell} 

\begin{abstract}
With the celebrated success of deep learning, some attempts to develop effective methods for detecting malicious PowerShell programs employ neural nets in a traditional natural language processing setup while others employ convolutional neural nets to detect obfuscated malicious commands at a character level.
While these representations may express salient PowerShell properties, our hypothesis is that tools from static program analysis will be more effective. 
We propose a hybrid approach combining traditional program analysis (in the form of abstract syntax trees) and deep learning. This poster presents preliminary results of a fundamental step in our approach: learning embeddings for nodes of PowerShell ASTs. We classify malicious scripts by family type and explore embedded program vector representations.
\end{abstract}



\maketitle

\section{Introduction}
\label{sec:intro}

PowerShell is a popular scripting language and a command-line shell. Originally only compatible with Windows, Powershell has gained a multitude of users over the last several years, especially with its cross-platform and open-source version, \emph{PowerShell Core}. PowerShell is built on the .NET framework and allows third-party users to write \textit{cmdlets} and \textit{scripts} that they can disseminate to others through PowerShell~\cite{pwsh-ms}. Along with increasing usage, PowerShell has also unfortunately been subject to malicious attacks through different types of computer viruses \cite{symantec-report}. PowerShell scripts can easily be encoded and obfuscated making it increasingly difficult to detect malicious activity \cite{hendler2018detecting}. According to the FireEye Dynamic Threat Intelligence (DTI) cloud, malicious PowerShell attacks have been rising throughout the past year~\cite{fireeye2018}. Detecting these malicious behaviors with Powershell can be challenging for a number of reasons. Attackers can perform malicious activity without deploying binaries on the attacked machines \cite{symantec-report}. Additionally, PowerShell is automatically downloaded on Windows machines. Further, attackers have shifted towards sophisticated obfuscation techniques that make detecting malicious scripts difficult \cite{paloalto17}. Notably, attackers use the \texttt{-EncodedCommand} flag to pass Base-64 encoded  commands bypassing the Powershell execution regulations. Recently, emerging research has deployed machine learning based models to detect malware in general~\cite{al2018adversarial,huang2018visual} and malicious PowerShell in particular~\cite{fireeye2018,hendler2018detecting}, where deep learning is employed to analyze malicious PowerShell scripts inspired by natural language understanding and computer vision approaches. Though these approaches may
support learning the features necessary to distinguish malicious scripts, with the wide range of obfuscation options used in Powershell scripts, we speculate that they might overlook some of the rich structural data in the codes. We therefore propose to break away from text-based deep learning and to use structure-based deep learning.
\begin{figure}[t]
	\includegraphics[width=0.49\textwidth]{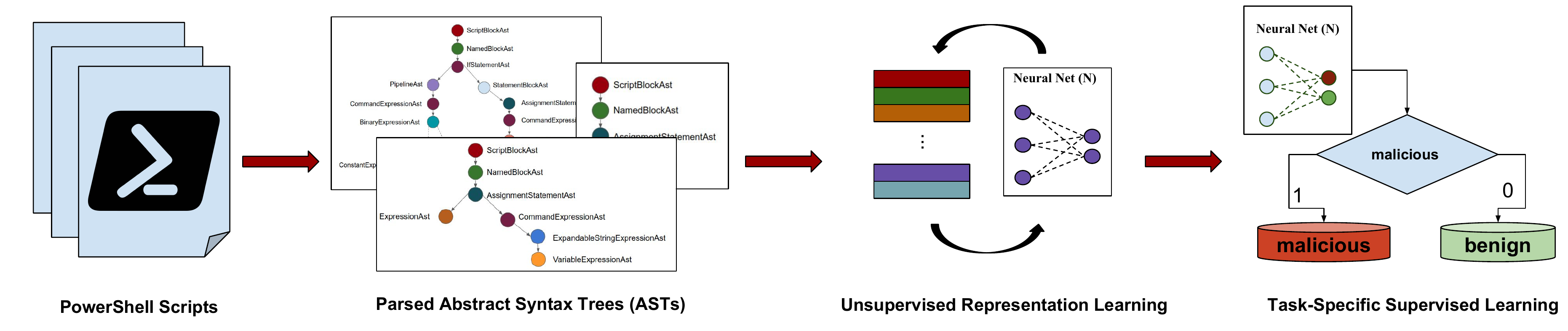}
	\caption{\small AST-based deep learning for malicious PowerShell detection.}
	\label{fig:framework}
\end{figure}
Our proposition is motivated by the successful use of Abstract Syntax Trees (ASTs) in manually crafting features to detect obfuscated PowerShell scripts~\cite{revoke-obfuscation}. While this use case does consider structural information, manually-crafted features can be vulnerable to high-level obfuscation (e.g., AST-based techniques~\cite{ast-obfuscation17}). Therefore, in this paper, we propose to learn representations of PowerShell scripts in an end-to-end deep learning framework based on their parsed ASTs. Specifically, we build on the work of \citet{peng2015building} to learn representations (embeddings) for AST nodes. These representations can then be incorporated in any of the tasks associated with PowerShell analysis, including malware detection as shown in Fig.~\ref{fig:framework}.


\section{Background}
\label{sec:background}
\paragraph{Deep Learning for PowerShell}
\citet{hendler2018detecting} proposed to use several deep learning models to distinguish benign and malicious PowerShell \emph{commands}. With a dataset of $6,290$ malicious and $60,098$ clean PowerShell commands, they implemented both Natural Language Processing (NLP) based detectors and detectors based on character-level Convolutional Neural Networks (CNNs) for text classification and treated the text as a raw signal at the character level. According to their results on different architectures (including a 9-layer CNN, a 4-layer CNN, and a long short-term memory net), all of the detectors obtained high AUC levels between $0.985$ and $0.990$. The authors suggest that the best performing classifier was an ensemble classifier that combined traditional NLP techniques with a CNN-based classifier. However, worse performance on their held out test set was observed with higher false positive rates. In a recent blog, FireEye~\cite{fireeye2018} apply a supervised classifier to detect malicious PowerShell commands leveraging a prefix-tree based stemmer for the PowerShell syntax. The input to the machine learning model is a vectorized representation of the stemmed tokens. The above propositions focused on detecting malicious PowerShell \emph{commands} rather than \emph{scripts} which are a more difficult challenge. Moreover, the features are derived from the commands' textual form, which may not capture the command's functional semantics and are prone to character frequency tampering.

\paragraph{AST for PowerShell} 
\citet{revoke-obfuscation} studied obfuscated PowerShell scripts. They presented a baseline character frequency analysis and used Cosine similarity to detect obfuscation in PowerShell scripts. They identify promising preliminary results and note a significant difference between obfuscated and non-obfuscated codes. Like \cite{hendler2018detecting}, the authors run into the issue of false negatives and suggest taking advantage of PowerShell Abstract Syntax Trees (ASTs) since PowerShell's API allows for simple AST extraction. Based on the parsed ASTs, the authors crafted $4098$ distributional features (e.g., distribution of AST types). The engineered feature vectors led to robust obfuscation classifiers on the test set. Similar to the character frequency tampering  challenge in text-based representations, the AST-based distributional features can be vulnerable to AST-based obfuscation~\cite{ast-obfuscation17}.

\paragraph{Deep Learning with AST}
\citet{peng2015building} developed a technique to build program vector representations, or embeddings, of different abstract syntax node types based on a corpus of ASTs for deep learning approaches.  They used nearest-neighbors similarity and k-means clustering to determine the accuracy of their resulting embeddings. They reported qualitative and quantitative results suggested that deep learning is a promising direction for program analysis. In this project, we build on~\cite{peng2015building}'s findings and further study this claim.


\section{Methods}
\label{sec:methods}

To learn a robust representation of PowerShell scripts, we take a hybrid approach combining traditional program analysis and deep learning approaches. We convert the PowerShell scripts to their AST counterparts, and then build embedding vector representations of each AST node type based on a corpus of PowerShell programs.

\paragraph{PowerShell scripts to Abstract Syntax Trees}
The considered dataset was composed of Base-64 encoded PowerShell scripts. Thus, as a preprocessing step, each PowerShell script/command was decoded. 
Given a decoded PowerShell script, we determined its abstract syntax tree representation by recursively traversing the script's properties using \texttt{[object.PSObject.Properties]} and storing  items  of type \texttt{[System.Management.Automation.Language.Ast]}. We stored the parent-child relationships among the AST nodes in a depth-first-search order as a text file. There were 37 different AST node types. With multi-core machines, ASTs generation can be carried out in parallel.

\paragraph{Preliminary Analysis of Abstract Syntax Trees}
After collecting the tree structures of our PowerShell scripts corpus, we conducted an exploratory analysis on the ASTs and their statistics. Furthermore, we used a random forest classifier to label a PowerShell script by its malware family type. As will be shown in~Section~\ref{sec:experiments}, few simple AST-based features were indicative of the malware family. 

\paragraph{Abstract Syntax Trees to Vector Representations} Having outlined our approach to the problem of malicious PowerShell programs, we herein take a fundamental step towards learning robust AST-based representations. We employed \cite[Algorithm 1]{peng2015building}
on the PowerShell dataset to learn real-valued vector representations of the $62$ AST node types. To this end, we parsed each constructed AST to a list of data structures to which we refer by \emph{subtrees}. A subtree of an AST represents a non-leaf node and its immediate child nodes, each labeled by its type. Next, we shuffled the subtrees to avoid reaching a local minima specific to a given script. For each subtree, with parent node $p$ and $n$ child nodes $\{c_i\}_{1\leq i \leq n}$, define $l_i = ({ \text{\# leaves of $c_i$} })/({\text{\# leaves of $p$}}). $ Similar to~\cite{peng2015building}, we define a loss function to measure how well the learnt vectors are describing the subtrees. Let $T$ be the number of distinct AST types whose embeddings we are trying to learn. Let $V \in \mathbb{R}^{ N_f \times T}$ be the embedding matrix of the AST node types and define $\vc(p)\subset V \in \mathbb{R}^{N_f \times 1}$ as the embedding vector that corresponds to the type of node $p$. The same holds for $\{\vc(c_i)\}_{1 \leq i \leq n}$. Additionally, let $W_l, W_r \in \mathbb{R}^{N_f \times N_f}$ be weight matrices and $b \in \mathbb{R}^{N_f \times 1}$ be a bias vector. Further, define $W_i$ as the weights matrix of node $i$ as

\begin{equation} W_i = \frac{n-i}{n-1}W_l + \frac{i-1}{n-1} W_r. \end{equation}
Let the distance metric $d$ be defined by
\begin{equation} d = ||\vc(p) - \tanh(\sum_{i=1}^n l_i W_i \cdot \vc(c_i) + b)||^2_2 .\end{equation}

Let $d_c$ be the distance function applied on a negative example of a given subtree where $k\leq n$ of the children nodes $\{c_i\}_{1 \leq i \leq n}$ are changed to different AST types. Given the parameters: $V, W_l, W_r, b$, we optimized $\max(0, \triangle + d - d_c)\;,$ the distance between a normal subtree's construction and that of a corrupted adversarial subtree. We used the Adam optimizer to find optimal embedding vectors and adjust the hyperparameters $\triangle$ and $k$. By default, $\triangle = 3, k = 3$. 

\section{Experiments}
\label{sec:experiments}

\paragraph{Setup.}
We utilize a corpus of hand-annotated and thoroughly analyzed malicious PowerShell scripts \cite{paloalto17}. 
This dataset consists of $4,079$ known malicious Powershell scripts annotated and classified based on their family types. These include ShellCode Inject, Powerfun Reverse, and others. The code repository will be made available at \url{https://github.com/ALFA-group}.

\begin{figure}
 \centering
 \includegraphics[width=0.3\textwidth, clip, trim=1.5cm 0.25cm 1.5cm 0cm]{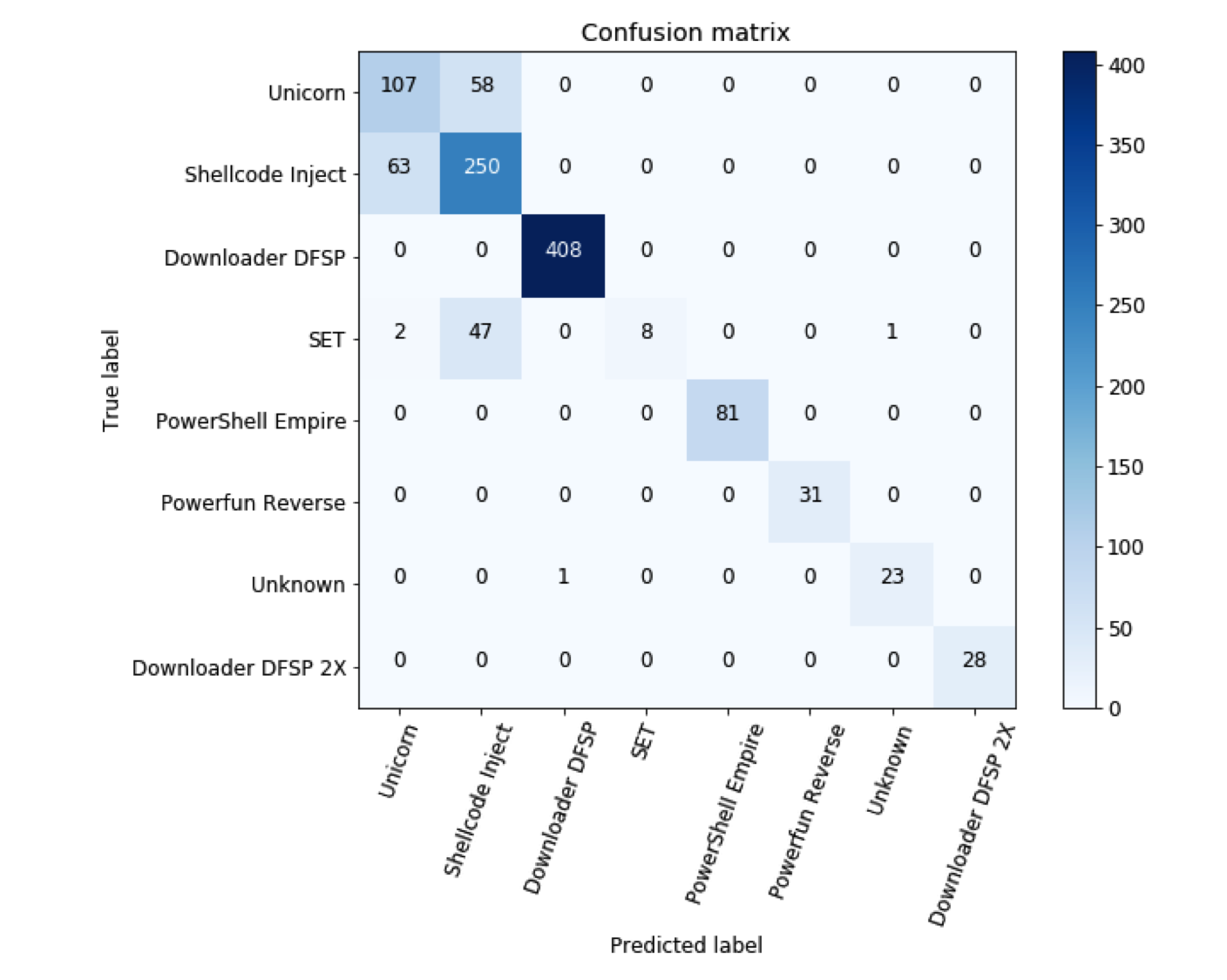}
 \caption{\small Heatmap for the confusion matrix results on the held out test set in the Malware Family Classification experiment.}
 \label{fig:family}
\end{figure}

\paragraph{Experiment 1: Malware Family Classification.} As a preliminary experiment, we attempted to classify malicious PowerShell scripts by family types. We used properties from the abstract syntax tree representation to conduct this classification. Specifically, we used only two features: depth and number of nodes per PowerShell AST. We used the family types as the labels of our classifier. Since the dataset used suffered from a class-imbalance problem, we weighted the classes when training the classifier (in this case a random forest classifier) based on how many examples each class contained. After hyperparameter tuning on maximum depth, we fit a classifier with a maximum depth of $11$. Due to sparsity of the dataset we used, we limited our experiment to family types with more than $40$ examples per family, resulting in eight different families. We randomly split the data into $70/30$ train/test split. The confusion matrix of the held-out test data is shown in Fig.~\ref{fig:family}. To our surprise, we found that two naive AST-based features---AST node count and AST depth---were enough to achieve an 3-fold cross-validation accuracy of $85\%$. Notably, even very simple features performed well because of the inherent program analysis background. This serves as a motivating example for the effectiveness of ASTs and exemplifies the power of harnessing ASTs to understand program representations.

\begin{figure}
 \centering
 \includegraphics[width=0.4\textwidth, clip, trim=0cm 0.75cm 0cm 0cm]{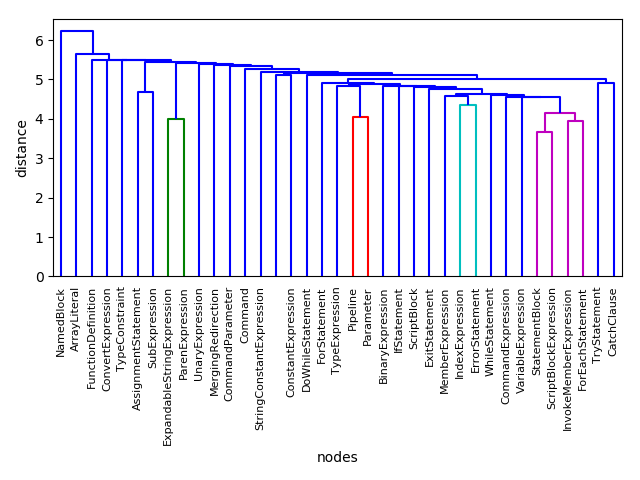}
 \caption{\small Dendrogram of node types and their relationships in the Learning Node Representations experiment.}
 \label{fig:dendrogram}
\end{figure}

\paragraph{Experiment 2: Learning AST Node Representations.} Extending these results, we build program vector representations of the dataset. As a case study, we analyzed a random sample of $10,000$ malicious subtrees from the total of $107,000$ subtrees in the malicious PowerShell corpus. This collection contained $37$ distinct AST node types comprising $175$ unique subtrees. We built the embedding matrix for these node types using the method described earlier. We trained our model for $200$ epochs until the loss stabilized towards 0. The qualitative results are summarized in a dendrogram in Fig.~\ref{fig:dendrogram}. It shows the relationships of embeddings with similar ones. Notably, the \texttt{TryStatement} and \texttt{CatchClause} node types are neighbors, as well as \texttt{ForStatement} and \texttt{DoWhileStatement}, and \texttt{Command} and \texttt{CommandParameter}. This is promising since one would expect such commands to serve similar functions in scripts. This preliminary experiment has limitations: for example, one would expect the \texttt{ForEachStatement} to land near the \texttt{ForStatement} as well. Additional training on the full malicious dataset is required to fully assess the validity of these methods.
As next steps, we hope to make use of these embeddings to build robust classifiers to classify a malicious script based on family. Afterwards, we will use these embeddings to build robust classifiers to determine if a given PowerShell script is malicious or not.

\section{Conclusion}
\label{sec:conclusion}

PowerShell scripts have targeted industries including Higher Education, High Tech, Professional and Legal Services, and Healthcare. This paper motivated the use of static program analysis (in the form of abstract syntax trees) to supplement deep learning techniques with rich structural information about the code, instead of text-based representations. We seek to use deep learning in an end-to-end unsupervised framework to identify intrinsic common patterns in our programs since even ASTs can be obfuscated. We saw that the depth and node count of an AST were enough to distinguish malware families and we took our first fundamental step in learning representations of PowerShell programs.

\section*{Acknowledgement}
\label{sec:acknowledgement}
\small 
This work was supported by the MIT-IBM Watson AI Lab and CSAIL CyberSecurity Initiative. We thank Palo Alto Networks for the dataset.

\tiny
\bibliographystyle{ACM-Reference-Format}
\bibliography{refs}

\end{document}